\documentclass[a4paper]{jpconf}
\usepackage{graphicx}
\begin{document}
\title{Magnetic Anisotropy in Single Crystalline CeAu$_{2}$In$_{4}$}
%\author{Jacky Mucklow}

%\address{Production Editor, \jpcs, \iopp, Dirac House, Temple Back, Bristol BS1~6BE, UK}

%\ead{jacky.mucklow@iop.org}
\author{Devang A. Joshi$^{1}$, P. Manfrinetti$^{2}$, S. K. Dhar$^{1}$ and A. Thamizhavel$^{1}$}

\address{$^{1}$Department of Condensed Matter Physics and Materials Science,
Tata Insitute of Fundamental Research, Dr. Homi Bhabha Road, Mumbai
400 005, India}

\address{$^{2}$Dipartimento di Chimica e Chimica Industriale, Università
di Genova, Via Deodecaneso 31, 16146 Genova, Italy}

\ead{devang@tifr.res.in}
\begin{abstract}
We have grown the single crystals of LaAu$_{2}$In$_{4}$ and CeAu$_{2}$In$_{4}$ by high temperature solution method and report on the anisotropic magnetic behavior of CeAu$_{2}$In$_{4}$ . The compounds crystallize in an orthorhombic structure with space group \textit {Pnma}.  LaAu$_{2}$In$_{4}$ shows a Pauli-paramagnetic behavior. CeAu$_{2}$In$_{4}$ do not order down to 1.8 K. The easy axis of magnetization for CeAu$_{2}$In$_{4}$ is along [010] direction. The magnetization data is analyzed on the basis of crystalline electric field (CEF) model. 
\end{abstract}

\section{Introduction}
Ce based compounds are known to exhibit a variety of interesting behavior
at low temperatures, such as normal trivalent, heavy fermion, mixed valent, Kondo lattice,
and quantum phase transition tuned by pressure, doping or magnetic field. The anomalous behavior is believed to
arise due to the proximity of 4$f$ energy levels to the Fermi level.
Single crystals are often used in such studies due to their inherently
superior crystalline order. Recently, studies on single crystals of orthorhombic
RAu$_{2}$In$_{4}$ (R = La, Ce, Pr and Nd) were reported by Salvador \textit{et al}. \cite{Salvador}. In their study, they found that 
the Ce ions in CeAu$_{2}$In$_{4}$ were in trivalent
state and do not show any signature
of magnetic ordering down to 2 K. This motivated us to undertake a
detailed study of this compound in the single crystalline form using
the techniques of magnetization, resistivity and heat capacity. Our
data reveal anisotropic magnetic behavior and indicate that the Ce
ions are in the normal trivalent state and may presumably order below 1.8~K. 

\section{Experiment}

%\begin{figure}[h]
%\begin{minipage}{36pc}
%\includegraphics[width=36pc]{Fig1.eps}
%\caption{\label{fig1}Top figure shows the photograph of CeAu$_{2}$In$_{4}$
%single crystal with the facing surface represent the {[}100{]} direction
%and {[}010{]} along the length. The bottom one represents the morphology
%of CeAu$_{2}$In$_{4}$ single crystal with the arrows indicating
%the corresponding direction.}
%\end{minipage}
%\end{figure}

We have grown the single crystals of RAu$_{2}$In$_{4}$ (R = La and Ce) by high temperature solution method using In as flux. The starting materials 3N-La and Ce, 5N-Au and 3N-In were taken in the stoichiometric amount with sample to flux
ratio of 1:19 in a high quality recrystallized alumina crucible. The crucibles were sealed in an evacuated quartz ampoule. The temperature of the furnace was raised to 1050 $^{\mathrm{o}}$C and dwelled there for 48 hrs in order to homogenize the melt. After homogenization, the temperature of the furnace was brought down to 500 $^{\mathrm{o}}$C over a period of 10 days and then rapidly cooled down to room temperature. The crystals were separated from the flux by centrifuging the melt in an evacuated glass tube at 300 $^{\mathrm{o}}$C. Small needle shaped crystals were obtained with lengths of 6 to 7 mm with width and height of order of 0.5 to 1 mm. The grown crystals were then oriented along the
%\begin{figure}[h]
%\includegraphics[width=21pc]{Fig1.eps}\hspace{1pc}%
%\begin{minipage}[b]{16pc}\caption{\label{fig1}Top figure shows the photograph %of CeAu$_{2}$In$_{4}$
%single crystal with the facing surface represent the {[}100{]} direction
%and {[}010{]} along the length. The bottom one represents the morphology
%of CeAu$_{2}$In$_{4}$ single crystal with the arrows indicating
%the corresponding direction.}
%\end{minipage}
%\end{figure}
principal crystallographic directions by means of Laue diffraction. The direction along the length of the crystal was found to the $b$ axis or [010] direction. The DC magnetic susceptibility and the magnetization measurements were performed in the temperature range 1.8 - 300 K and in the magnetic fields up to 70 kOe along the three principal directions using a Quantum Design SQUID magnetometer, high field magnetization measurements up to a field of 120 kOe were performed using a vibrating sample magnetometer (VSM, Oxford Instruments). The heat capacity measurements were carried out using physical property measurement system (PPMS, Quantum Design) and the resistivity measurements were performed on a  home built setup.

\section{Results and Discussion}

\begin{figure}[h]
\begin{minipage}{18pc}
\includegraphics[width=18pc]{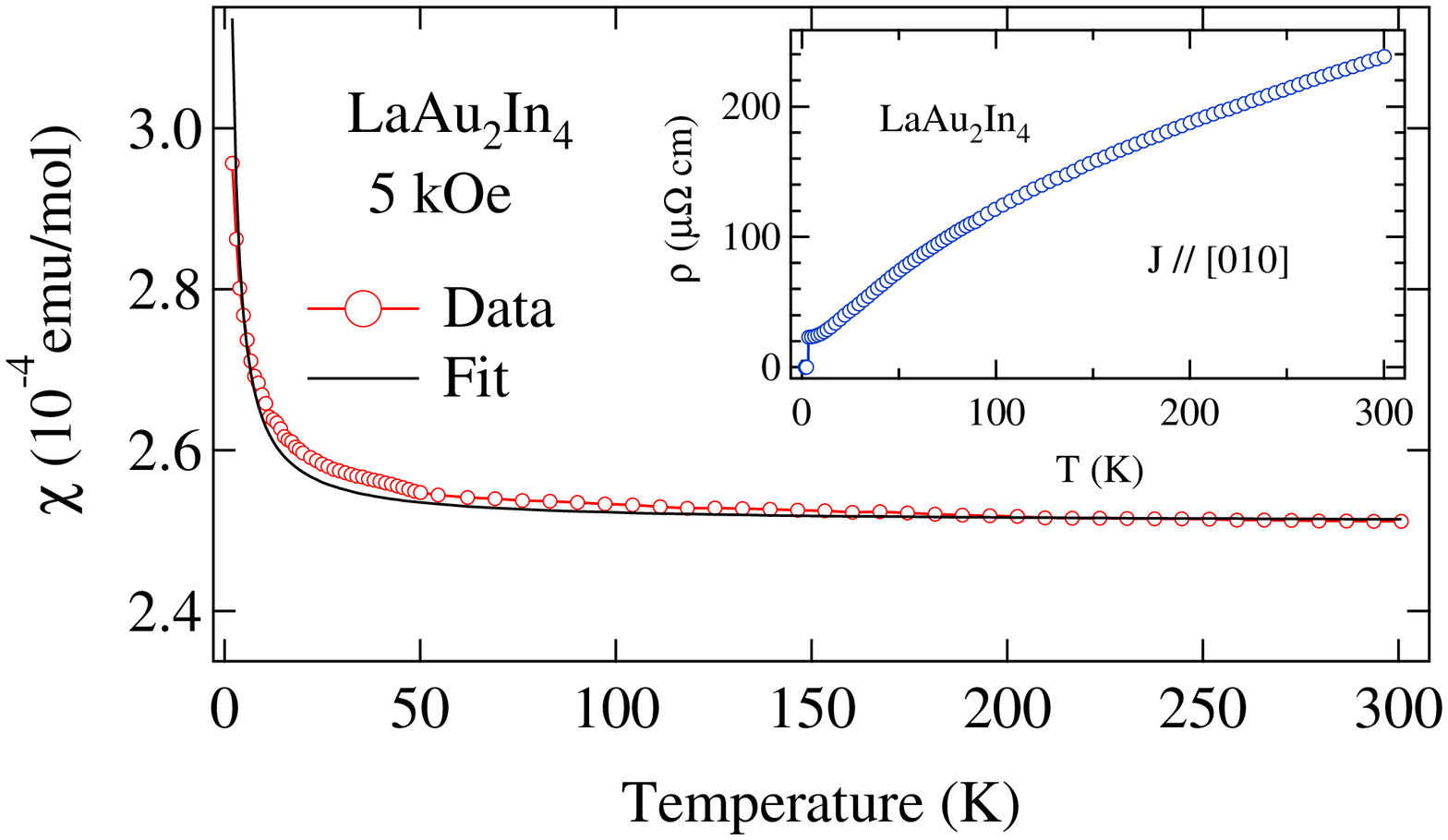}
\caption{\label{fig:La}Susceptibility of LaAu$_{2}$In$_{4}$ with modified
Curie-Weiss fit. The inset shows its resistivity with current parallel to [010] direction.}
\end{minipage}\hspace{2pc}%
\begin{minipage}{18pc}
\includegraphics[width=18pc]{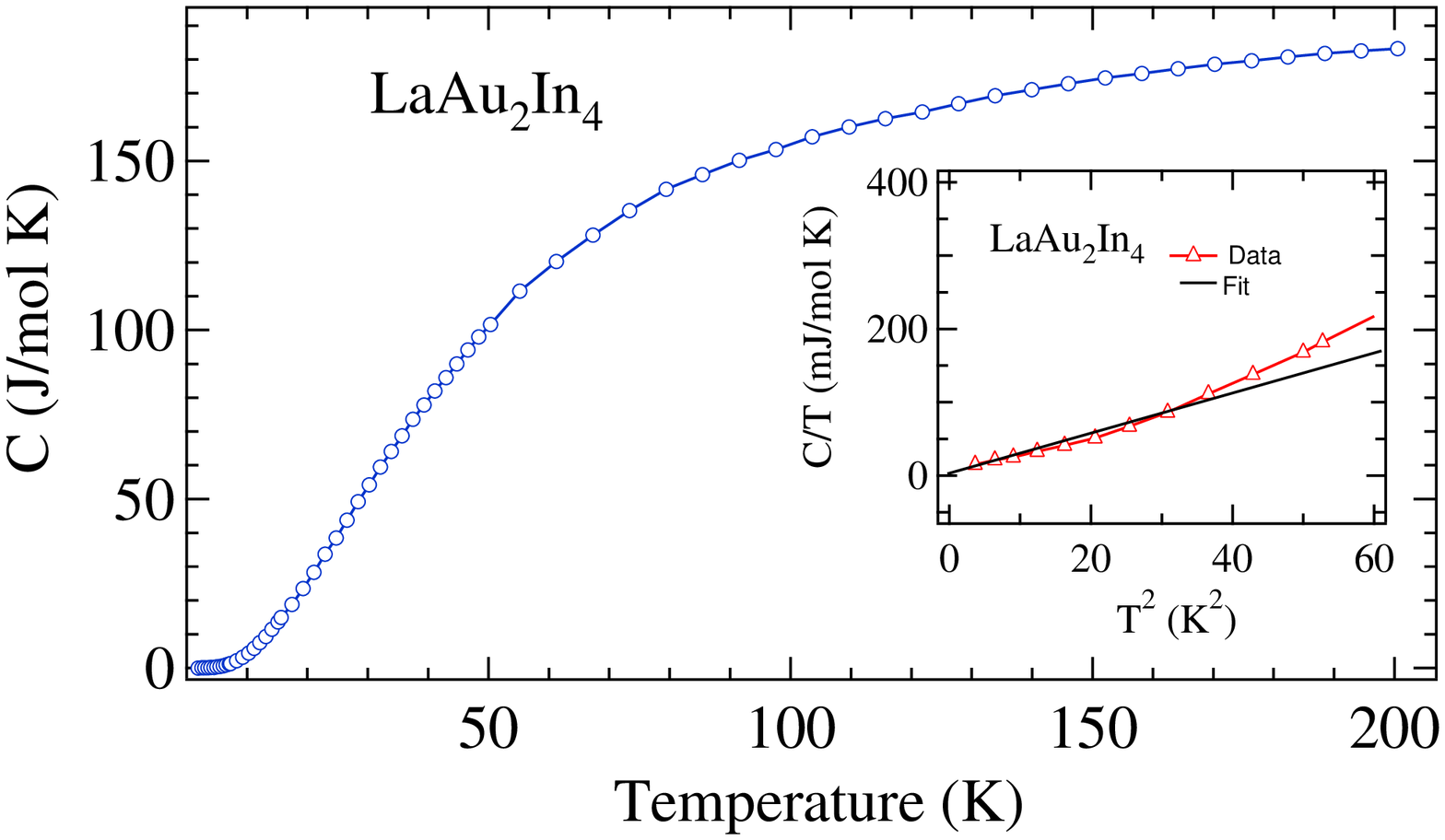}
\caption{\label{LaHC}Heat capacity of LaAu$_{2}$In$_{4}$ with the inset showing the low temperature C/T \textit{vs} T$^{2}$ curve.}
\end{minipage}
\end{figure}

RAu$_{2}$In$_{4}$ (R = La and Ce) compounds crystallize in an orthorhombic
structure with a space group $Pnma$. In order to confirm the phase homogeneity of the compound and lattice parameters, a Rietveld analysis of the observed x-ray pattern of both the compounds was done using the FullProf program. The lattice parameters thus obtained are $a$ = 18.51~\r{A}, $b$ = 4.685~\r{A}, and $c$ = 7.339~\r{A} for LaAu$_{2}$In$_{4}$ and $a$ = 18.505~\r{A}, $b$ = 4.667~\r{A} and $c$ = 7.318~\r{A} for CeAu$_{2}$In$_{4}$. The unit cell volumes are 636.4 and 632~\r{A}$^{3}$ respectively for La and Ce compounds as expected on the basis of lanthanide contraction, but the reported unit cell volume~\cite{Salvador} of Ce compound is higher than that of La compound. 

LaAu$_{2}$In$_{4}$ shows a Pauli-paramagnetic behavior (Fig. \ref{fig:La}) at
high temperature with an upturn at low temperature. The curve was fitted to modified Curie-Weiss law with an effective moment of 0.1 $\mu_{B}$. The low value of effective moment indicates that the upturn arises from some paramagnetic impurity (trace quantities of other rare earths in the La used) in the sample. The value obtained for $\chi_{0}$ is $2.52\times10^{-4}$, which is typical for the non-magnetic La compounds and is close to the reported one. The resistivity of the compound with J // [010] (inset of Fig. \ref{fig:La}) decreases with temperature showing a metallic behavior down to 3.7~K followed by a superconducting drop due to the presence of  free In on the sample surface. The resistivity measurement could only be done with J~$\parallel$~[010] direction (length of the crystal), because of the small dimensions of crystal along the other directions. The resistivity is not linear as expected for a non magnetic compound, but shows a positive curvature. Such a positive curvature is attributed to $s-d$ inter-band scattering of the conduction electrons as described by Mott and Jones  \cite{Mott}. The heat capacity as shown in Fig.~\ref{LaHC} is typical for the non-magnetic compound. The data could not be described by Debye, Einstein, or a combined model. The low temperature C/T \textit{vs} T$^{2}$ curve (inset of Fig. \ref{LaHC}) reveals $\gamma=5$ mJ/mole K$^{2}$.

\begin{figure}[h]
\begin{minipage}{18pc}
\includegraphics[width=18pc]{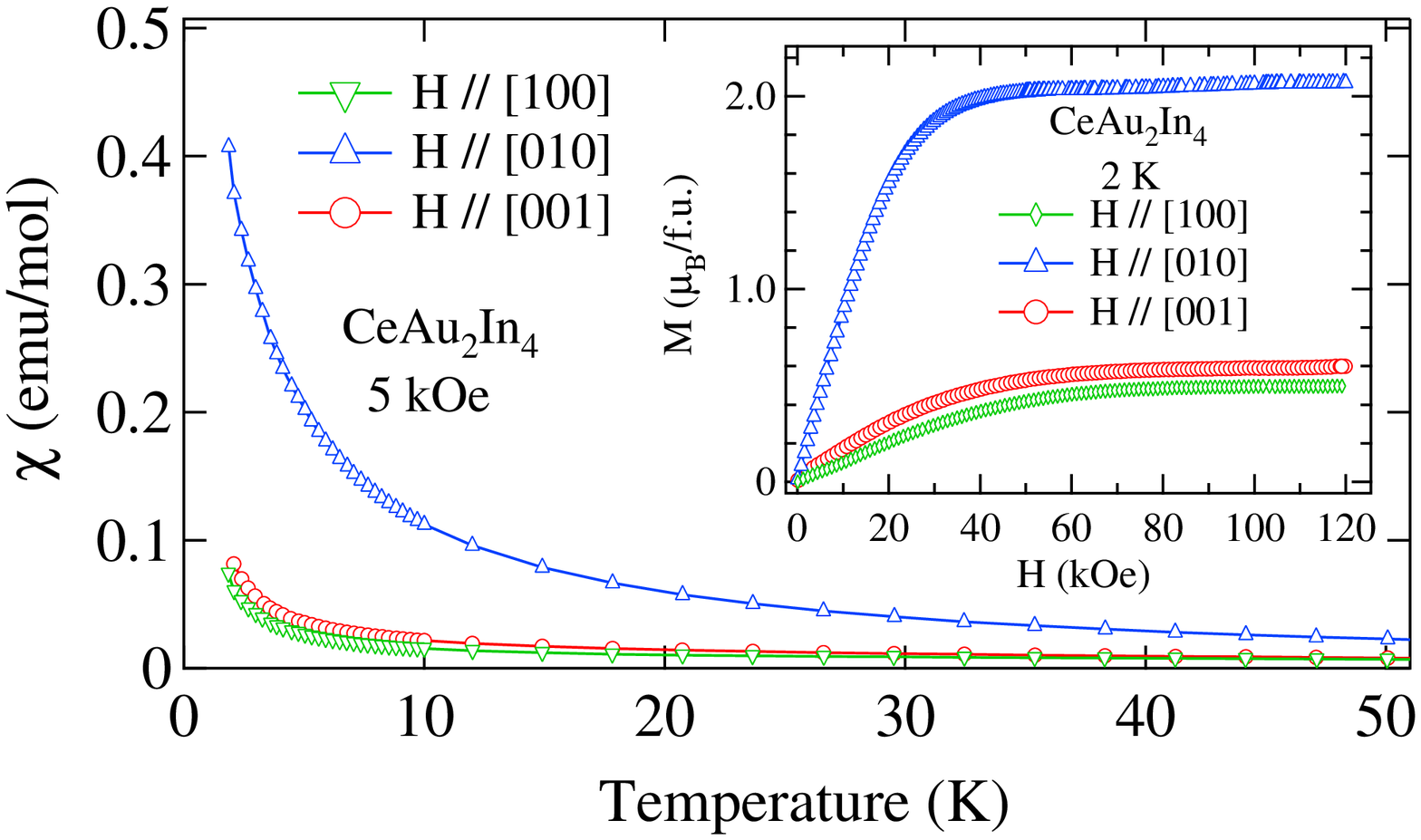}
\caption{\label{Mag_Ce}Magnetic susceptibility of CeAu$_{2}$In$_{4}$
with field parellel to the three crystallographic directions. The
inset shows the magnetic isotherm at 1.8 K with field parallel to indicated
directions.}
\end{minipage}\hspace{2pc}%
\begin{minipage}{18pc}
\includegraphics[width=18pc]{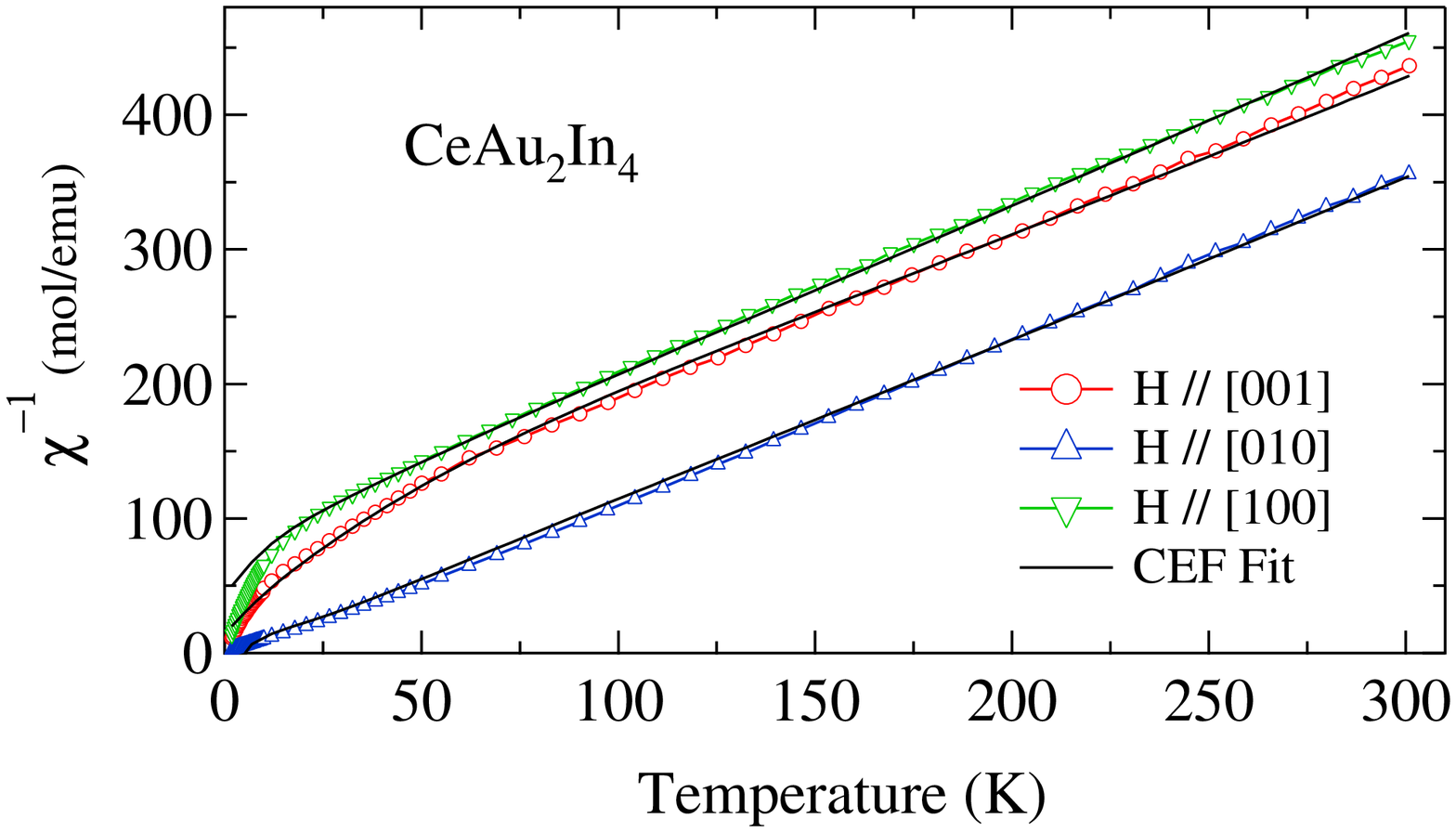}
\caption{\label{CEF}Inverse magnetic susceptibility of CeAu$_{2}$In$_{4}$ with with field parallel to the indicated directions and crystal electric field fit described in text.\hspace{ 5 in}~} 
\end{minipage}
\end{figure}

The magnetic susceptibility of CeAu$_{2}$In$_{4}$ is shown in Fig.
\ref{Mag_Ce}. There is no signature of magnetic ordering down to 1.8 K. The susceptibility with field along {[}010{]} direction is higher compared to that along the other two directions, indicating the easy axis of magnetization. The inverse susceptibility of the compound was fitted to modified
Curie-Weiss law (not shown). The effective moments obtained are 2.55,
2.55 and 2.54 $\mu_{B}$ respectively for field parallel to {[}100{]},
{[}010{]} and {[}001{]} direction. The paramagnetic Curie temperatures
are -70, 10 and -54 K and $\chi_{0}$ of $2.01\times10^{-5}$, $5.2\times10^{-5}$
and $3.7\times10^{-5}$ emu/mol respectively. The effective moments
are close to that expected for the Ce$^{3+}$ ion. The polycrystalline
average of the paramagnetic Curie temperature is -38 K, indicating
an antiferromagnetic type of interaction among the Ce moments.

The magnetic isotherm of the compound at 2 K is shown as an inset
of Fig. \ref{Mag_Ce}. The magnetization for field along {[}010{]}
increases sharply with field and saturates above $\approx$ 50 kOe.
The saturation moment obtained is 2.07 $\mu_{B}\mathrm{/Ce}$, close
to the saturation moment of the Ce$^{3+}$ ion (2.14 $\mu_{B}\mathrm{/Ce}$),
as expected for the easy axis of magnetization. The magnetization
along {[}100{]} and {[}001{]} is $\approx$ 0.5 and 0.6 $\mu_{B}\mathrm{/Ce}$
respectively, much less then the saturation moment of the Ce$^{3+}$
ion indicating the hard axes of magnetization. The behavior of magnetic
isotherms is similar to a ferromagnetically ordered compound. This indicates that the energy gained due to the field induced ferromagnetic alignment of the moments along the easy axis of magnetization exceeds the thermal and zero field mutual interaction between the moments. The non ordering of the compound down to 1.8~K is not due to the crystal field or Kondo effect but may be due to a weak exchange interaction between the Ce ions. It may be noticed that the shortest Ce-Ce distance is 4.66~\r{A} along the \textit {b} axis while it is even longer along the other two directions. It is to be noted here that both Pr and Nd compounds also do not order down to 2~K \cite{Salvador}.

To estimate the effect of crystal fields in CeAu$_{2}$In$_{4}$  we fitted the inverse susceptibility using the point charge crystal
field model. The Ce atom in the orthorhombic CeAu$_{2}$In$_{4}$
occupies the 4\textit{c }Wyckoff position and  possesses a monoclinic
site symmetry, but for the simplicity of crystal field calculations
we used CEF Hamiltonian for the orthorhombic site symmetry. The Hamiltonian
for the orthorhombic site symmetry is given by 

\begin{equation}
\mathcal{H}_{{\rm CEF}}=B_{2}^{0}O_{2}^{0}+B_{2}^{2}O_{2}^{2}+B_{4}^{0}O_{4}^{0}+B_{4}^{2}O_{4}^{2}+B_{4}^{4}O_{4}^{4} + B_{6}^{0}O_{6}^{0} + B_{6}^{2}O_{6}^{2} + B_{6}^{4}O_{6}^{4} + B_{6}^{6}O_{6}^{6},\end{equation}
where $B_{\ell}^{m}$ and $O_{\ell}^{m}$ are the CEF parameters and
the Stevens operators, respectively~\cite{Hutchings,Stevens}. For Ce atom the $O_6$ Stevens operator is zero and hence the last four terms in the Hamiltonian are zero. The inverse susceptibility with field along the
three crystallographic axes was fitted to the above CEF model (Fig. \ref{CEF}) with the unique value of parameters given by $B_{2}^{0}=2$ K, $B_{2}^{2}=2$
K, $B_{4}^{0}=-0.8$ K, $B_{4}^{2}=-1$ K and $B_{4}^{4}=6.5$ K and
the molecular field exchange constant $\lambda_{100}=-55$ , $\lambda_{010}=25$
and $\lambda_{001}=-13.4$ mol/emu. The negative value of the exchange
constant indicates the overall antiferromagnetic interaction between
the moments. A positive value along the {[}010{]} direction indicates
a ferromagnetic type of interaction along that direction. But ferromagnetically
aligned moments may couple antiferromagnetically giving an overall
antiferromagnetic type of behavior. The energy levels estimated from
the CEF parameters are three doublets E$_{0}$= 0, E$_{1}$= 73 K and
E$_{2}$= 412 K.

\begin{figure}[h]
\begin{minipage}{18pc}
\includegraphics[width=18pc]{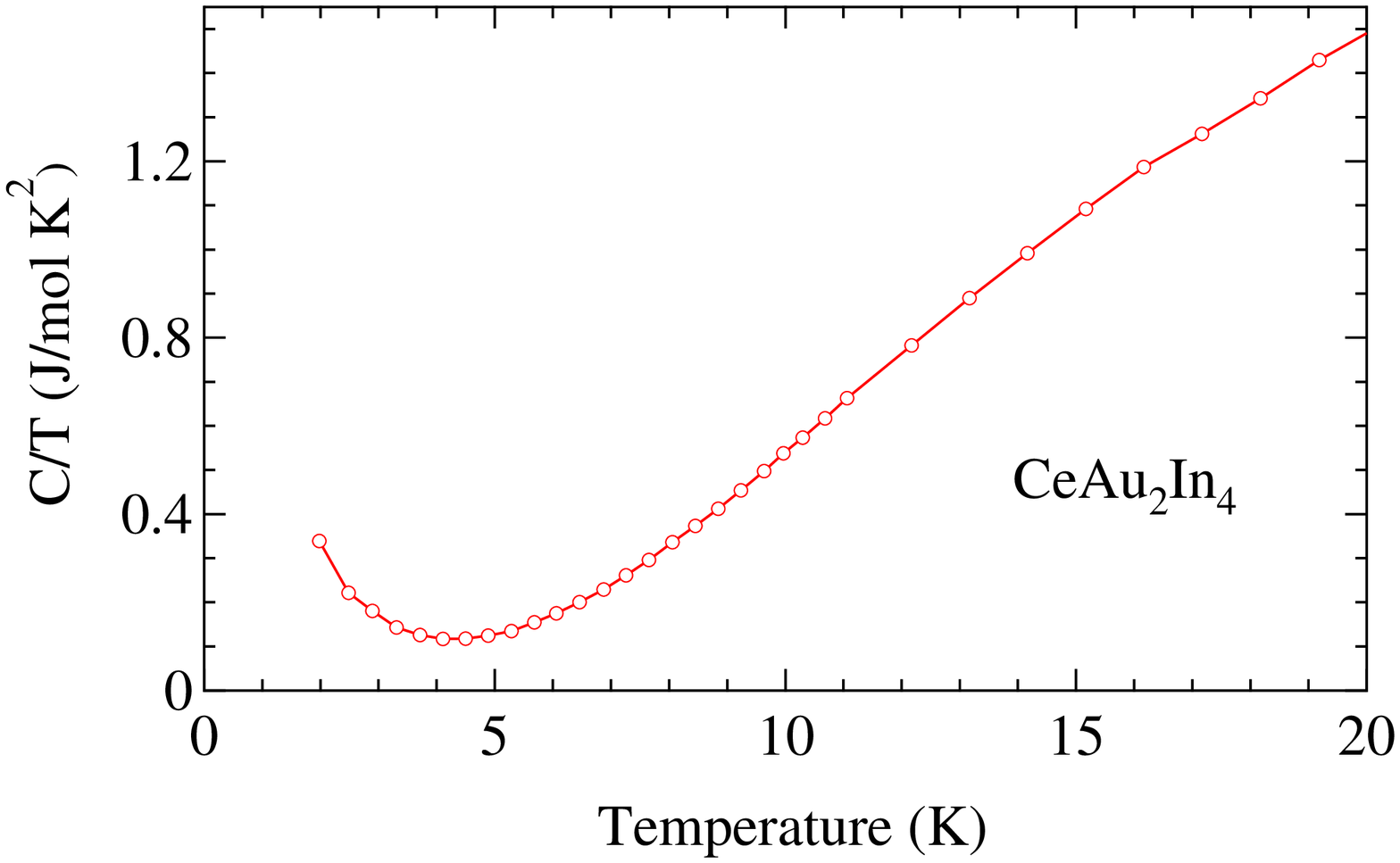}
\caption{\label{HC_Ce}a) Low temperature C/T \textit{vs} T curve for CeAu$_{2}$In$_{4}$.}
\end{minipage}\hspace{2pc}%
\begin{minipage}{18pc}
\includegraphics[width=18pc]{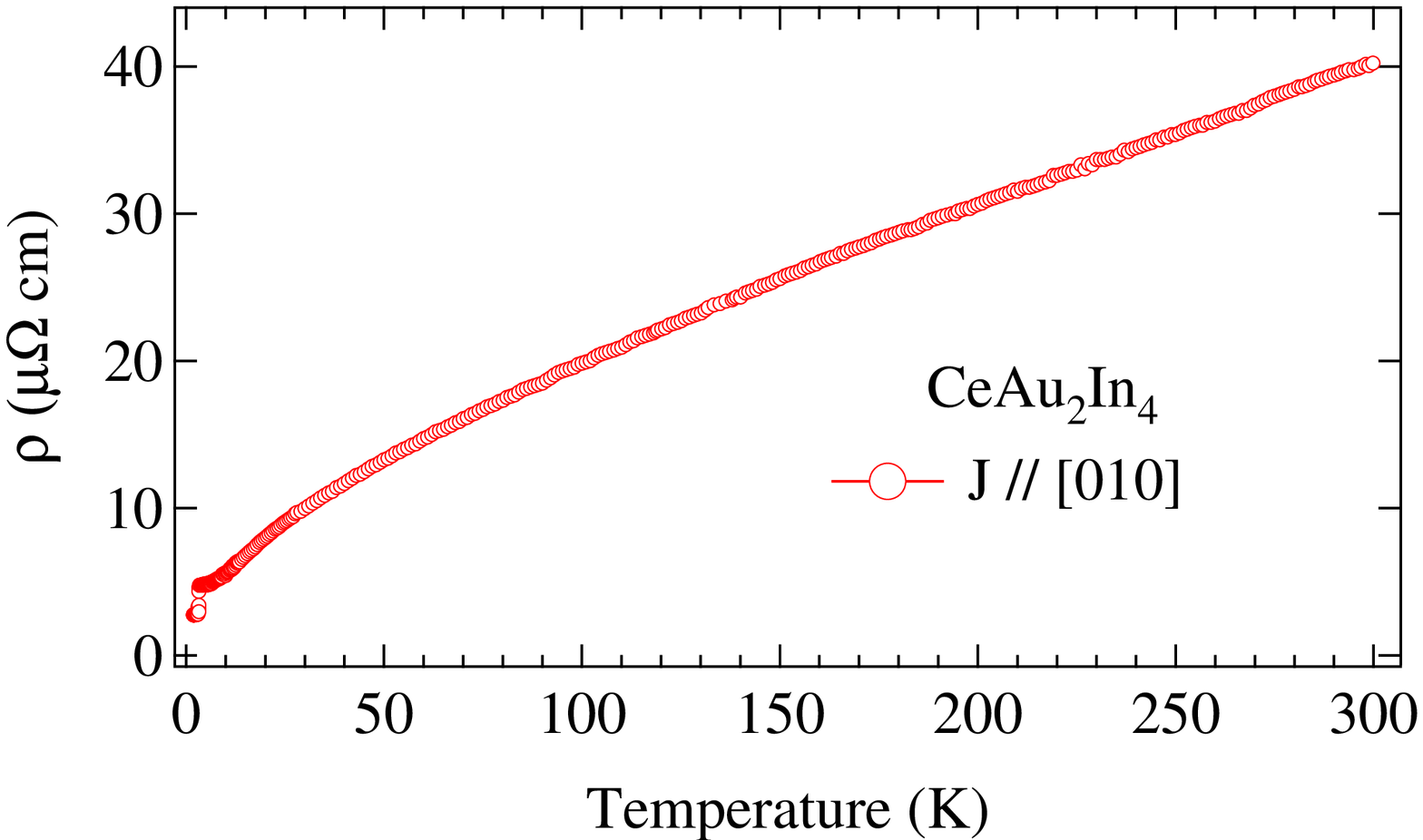}
\caption{\label{ResCe}Resistivity curve of CeAu$_{2}$In$_{4}$ with the inset showing the low temperature part.}
\end{minipage}
\end{figure}

 The heat capacity of CeAu$_{2}$In$_{4}$ in the form of C/T vs T$^{2}$
(Fig. \ref{HC_Ce}) shows an upward turn below $\approx$ 4~K, which
may be due to the onset of magnetic ordering. The temperature dependence of electrical resistivity from 1.8-300~K is shown in Fig.~\ref{ResCe}. The resistivity shows a typical metallic behavior down to 3.7~K followed by a drop at low temperature. The drop at 3.7~K is due to the superconductivity of traces of free In present on the surface of the crystal. The resistivity shows a curvature around 100 K, which may be attributed to crystalline electric field effect. The electrical resistivity does not exhibit features accounted with Kondo behavior, indicating that the hybridization between the  conduction electrons and 4$f$ orbitals is negligible.

\section{Summary}

To summarise we have grown the single crystals of LaAu$_{2}$In$_{4}$ and CeAu$_{2}$In$_{4}$ by high temperature flux method. We do not see any superconducting transition in LaAu$_{2}$In$_{4}$ down to 1.8~K. Ce ions in CeAu$_{2}$In$_{4}$ exhibit normal trivalent behavior with significant anisotropy. The exchange interaction is weak, presumably due to large Ce-Ce distance and the Ce ions remain in a paramagnetic state down to 2~K. The {[}010{]} direction was found to be the easy axis of magnetization and CEF analysis shows an overall antiferromagnetic interaction between the Ce ions.

\section{Acknowledgment}
We thank Dr. R. Nagalakshmi for her kind help in carrying out the work.


\section*{References}
\begin{thebibliography}{9}
\bibitem{Salvador} James R. Salvador, et al 2007 \textit{Inorg. Chem.} \textbf{46} 6933.
\bibitem{Mott} N. F. Mott and H. Jones 1958 \textit{The Theory of the Properties of Metals and Alloys} (London: Oxford Uni. Press).
\bibitem{Hutchings} M. T. Hutchings 1965 \textit{Solid State Physics: Advances in Research and Applications} \textbf{Vol.16} edited by F. Seitz and B. Turnbull (New York: Academic) p.227.

\bibitem{Stevens} K. W. H. Stevens 1952 \textit{Proc. Phys. Soc.} (London)  \textbf{Sect.A65} 209.

\end{thebibliography}
\end{document}